\documentclass[a4paper,fleqn,usenatbib]{mnras}

\usepackage[T1]{fontenc}
\usepackage{ae,aecompl}

\usepackage{graphicx}	
\usepackage{amsmath}	
\usepackage{amssymb}	

\usepackage{newtxtext,newtxmath}

\newcommand{\equ}[1]{eq.~(\ref{eq:#1})}

\newcommand{\fig}[1]{Fig.~\ref{fig:#1}}

\newcommand{\be}{\begin{equation}}
\newcommand{\ee}{\end{equation}}
\newcommand{\bea}{\begin{eqnarray}}
\newcommand{\eea}{\end{eqnarray}}

\newcommand{\ifm}[1]{\relax\ifmmode#1\else$\mathsurround=0pt #1$\fi}
\newcommand{\kms}{\ifmmode\,{\rm km}\,{\rm s}^{-1}\else km$\,$s$^{-1}$\fi}

\newcommand{\ltsima}{$\; \buildrel < \over \sim \;$}
\newcommand{\lsim}{\lower.5ex\hbox{\ltsima}}
\newcommand{\gtsima}{$\; \buildrel > \over \sim \;$}
\newcommand{\gsim}{\lower.5ex\hbox{\gtsima}}

\usepackage{color}

\title[Dust Attenuation of FirstLight galaxies at high-$z$]{Dust attenuation in galaxies at cosmic dawn from the FirstLight simulations}

\author[M. Mushtaq et al.]{Muzammil Mushtaq,$^{1}$ Daniel Ceverino, $^{2,3}$\thanks{E-mail: daniel.ceverino@uam.es} Ralf S. Klessen, $^{1,4}$ Stefan Reissl $^{1}$ and
\newauthor
 Prajwal Hassan Puttasiddappa$^{5,6}$
\\
$^{1}$ Zentrum für Astronomie, Ruprecht-Karls-Universität Heidelberg, Albert-Ueberle-Strasse 2, 69120 Heidelberg, Germany\\
$^{2}$Departamento de Fisica Teorica, Modulo 8, Facultad de Ciencias, Universidad Autonoma de Madrid, 28049 Madrid, Spain\\
$^{3}$CIAFF, Facultad de Ciencias, Universidad Autonoma de Madrid, 28049 Madrid, Spain \\
$^{4}$Universit\"{a}t Heidelberg, Interdisziplin\"{a}res Zentrum f\"{u}r Wissenschaftliches Rechnen, INF 205, 69120, Heidelberg, Germany\\
$^{5}$Institut für Theoretische Physik, Ruprecht-Karls-Universität Heidelberg, Philosophenweg 16, 69120 Heidelberg, Germany\\
$^{6}$PPGCosmo, Universidade Federal do Espírito Santo, 29075-910, Vitória, ES, Brazil
}
\date{Accepted XXX. Received YYY; in original form ZZZ}

\pubyear{2023}

\begin{document}
\label{firstpage}
\pagerange{\pageref{firstpage}--\pageref{lastpage}}
\maketitle

\begin{abstract}
We study the behavior of dust in galaxies at cosmic dawn, $z=6-8$, by coupling the FirstLight simulations with the radiative transfer code POLARIS. 
The starburst nature of these galaxies and their complex  distribution of dust lead to a large diversity of attenuation curves.
These follow the Calzetti model only for relatively massive galaxies, M$_{\rm stars}\simeq 10^9 M_{\odot}$. Galaxies with lower masses have steeper curves, consistent with the model for the Small Magellanic Cloud (SMC). 
The ultraviolet and optical slopes of the attenuation curves are closer to the modified Calzetti model, with a slight preference for the power-law model for galaxies with the highest values of attenuation.
We have also examined the relation between the slope in the far-ultraviolet, $\beta_{UV}$, and the infrared excess, \textit{IRX}.
 At $z=6$, it follows the Calzetti model with a shift to slightly lower $\beta_{UV}$ values due to lower metallicities at lower attenuation.
 The same relation at $z=8$ shows a shift to higher \textit{IRX} values due to a stronger CMB radiation at high-$z$.
\end{abstract}



\section{Introduction}

Galaxies at cosmic dawn generally refer to early galaxies starting from redshift $z\simeq20$, where we expect their formation, till the end of the reionization era at $z\simeq5$. 
In the cosmological context, the first galaxies are the result of hierarchical structure formation in the early Universe. It starts with the formation of the first stars with extremely low metallicities \citep{Bromm:2001bi, metalpoor1, 2012, Yoshida, Glover2012TheFS, 2019, KlessenGlover23}.  
As galaxies evolve, the metal content within their interstellar medium (ISM) increases as a consequence of the ejection of metals from supernova explosions \citep{supernove1, supernove2} and stellar winds. This injection of metals into the ISM is also modulated by galactic-scale inflows and outflows  \citep{stellarwind, galacticwind}.
Eventually dust forms and grows in dense and metal-rich regions of clouds, supernova shells, or stellar winds \citep{dustformationreview}. 
Therefore, dust is usually thought to be proportional to the amount of metals in the ISM \citep{2013}. 

The presence of pervasive dust in high-redshift galaxies is observed through two effects: (a) dust grains are responsible for the scattering and absorption of high energy photons coming from stars and active galactic nuclei (AGN)
at ultraviolet (UV) and optical wavelengths,  (b) and the grains re-emit the absorbed energy into the infrared (IR) region. 
Although the IR emission from AGN is significant in some high-$z$ galaxies \citep{Labbe23}, this paper focuses on the interaction between starlight and dust.

This complex interaction is usually encapsulated in the process of dust attenuation \citep{2020salim}. 
The most widely used attenuation curve for high-redshift studies is the empirical law derived by \citet{1994ApJ...429..582C} from a sample of 39 nearby starburst galaxies. 
The model distinguishes between the attenuation of a single point source star (foreground dust), and the attenuation of an extended object such as a starburst galaxy seen through a foreground dusty medium.
This method uses Balmer lines $(\text{H}_{\alpha}/\text{H}_{\beta})$ to characterize the dust extinction. The UV/optical spectra of sampled galaxies were split based on the values of their Balmer optical depth and the smallest Balmer optical depth is considered as the reference spectrum against which the rest of the spectra are compared. This relative attenuation is then anchored with the IR data to get an absolute curve \citep{2000calzetti}.

The relation between the slope of the rest-frame UV continuum, $\beta_{UV}$, and the infrared excess, \textit{IRX}, is considered one of the most important outputs to examine dust attenuation in galaxies.
The rest-frame UV continuum flux is expressed as a power law function of wavelength, $f_{\lambda} \propto \lambda^{\beta}$, between $\lambda=1000$ and 3000 \AA. 
The infrared excess is the ratio of the infrared to the UV luminosity, $IRX=L_{IR}/L_{UV}$, where $L_{IR}$ is the total IR luminosity in the range $10^{4}$ to $10^{7}$\AA, and $L_{UV}$ is the UV luminosity at 1500 \AA \ \citep{2017narayan}.
The UV continuum slope is sensitive to the presence of a young stellar population with low metallicities. This drives very low values, $\beta_{UV}<-2$. On the other hand, the presence of significants amounts of dust reddens the stellar spectrum and increases $\beta_{UV}$ accordingly. 

The IR emission is the result of two phenomena: (a) The UV light from stars (or AGN) heats the dust grains, which emit IR light. (b) For high-$z$, the CMB has a stronger intensity penetrating into the ISM and it can also heat the grains. Dust in low-mass galaxies, in which it is not sufficiently heated by the stellar UV light, can also be heated by the CMB. 
As the CMB temperature decreases with redshift this heating process contributes more to the IR emission at $z = 8$ than at $z\simeq 6$ \citep{2013cunha}, for fixed stellar mass.
A wide range of theories is available to understand the origin of the $IRX-\beta_{UV}$ relation.
In general, increasing the dust optical depth moves galaxies towards a more positive UV slope with an increase in \textit{IRX}. On the other hand, more complex dust-star geometries can vary the attenuation curves, resulting in changes in \textit{IRX} \citep{2021_liang}.

Understanding the role of dust attenuation in galaxies at high-$z$ is challenging due to the lack of detailed observations. 
It is often assumed attenuation curves derived from low-$z$ observations \citep{2000calzetti}.
It is unclear whether these relations hold in the dense and low-metallicity conditions in the ISM of galaxies at cosmic dawn. 
As an alternative path, cosmological simulations act as convenient tools for making mock observations that could be compared with future measurements. 
Through them, we can study different properties of galaxies, their evolution, and the thermal behavior of dust grains using radiative transfer (RT) simulators \citep[e.g.][]{2017, 2018, 2019dust}.

In this paper, we study the dust attenuation in galaxies from the FirstLight (FL) simulations \footnote{\href{https://www.ita.uni-heidelberg.de/~ceverino/FirstLight/}{www.ita.uni-heidelberg.de/$\sim$ceverino/FirstLight/}} \citep{Ceverino17b}. 
Large-box cosmological simulations, which usually produce a large population of galaxies, cannot resolve their internal properties, such as a multi-phase ISM \citep[e.g.][]{2014MNRAS.445..175G, 10.1093/mnras/stw2869}. On the other hand, traditional zoom-in cosmological simulations have high resolutions but only contain a small number of galaxies. Therefore, their conclusions typically cannot be extrapolated to the general galaxy population  \citep[e.g.][]{10.1093/mnras/stv1679, 10.1093/mnras/stx1792, 10.1093/mnras/stx335}. 
The FirstLight simulations contain a volume and mass complete sample of galaxies, resolved with parsec-scale resolution. 
A series of FL papers have been published, testing the physical properties of galaxies at cosmic dawn against observations \citep{2018DC, 2019DC, 2020langan, Ceverino21}. 

Our objective is to study the attenuation of the stellar light by dust in the FL simulated galaxies at redshifts $z = 6$ and $8$. So we first extract star-particle and dust density distributions from FL as an input for an (RT) simulator. 
The idea behind this approach is to compute the propagation of light from stellar sources through the dusty ISM environment. We use the POLARIS\footnote{\href{https://portia.astrophysik.uni-kiel.de/polaris/}{portia.astrophysik.uni-kiel.de/polaris/}} code \citep{polaris} that is based on the 3D Monte-Carlo (MC) technique. POLARIS is able to compute the direct, scattered, and re-emitted light from the sources (i.e., star-particles). The outcome of the POLARIS simulation is the transmitted light in UV/optical wavelengths and the emission from dust in the infrared. In this work, we focus on two tasks:  
The first goal is to calculate the attenuation curves for sampled FL galaxies at $z = 6$ and 8. This topic has been explored through observations of nearby galaxies and also using RT simulators with various dust intrinsic models and geometrical distribution of dust and stars \citep{Calzetti_2001, Narayanan_2018, 2020salim}. The steepness of the attenuation curve in UV and optical bands is a key feature that shows considerable variability in the observed and simulated results \citep{AlvarezMarquez19, 2021_liang}.  This is an indication of the complex mixture of dust and stars in the interstellar medium of high-$z$ galaxies. We also compare attenuation curves of FL galaxies with the existing models and ask how their slopes depend on the physical parameters of galaxies e.g, stellar masses.
The second goal is to study the IRX-$\beta_{UV}$ relation and its shape at high-$z$. This takes into account the dust attenuation of UV light and its re-emission in the infrared. A number of interesting aspects arise from a potential offset of such relation from the canonical models. For example, bright and dusty star-forming galaxies at $z \geq 2$ tend to exhibit higher infrared luminosity along with  bluer UV slopes \citep[e.g.][]{2017bourne, 2010ApJ...712.1070R, Casey_2014, AlvarezMarquez19}. 

This paper is organized as follows. In Section~\ref{sec:methodology}, we describe the techniques and selection criteria for galaxies and star-particles that are assembled into POLARIS. In Section~\ref{sec:resultsdis} we present the results and discussion. This section is split into two parts, (a) the behavior of attenuation curves, and (b) the diversity of the $IRX-\beta_{UV}$ relation. Finally, we summarize our results in Section~\ref{sec:conclusion}. 

\section{METHODOLOGY}\label{sec:methodology}

\subsection{FirstLight simulations}

 The FirstLight simulations \citep{Ceverino17b} are multi-object, zoom-in cosmological simulations within three different cosmological boxes with sides 10, 20, and 40 Mpc/h.
 The simulations are performed with the $N$-body + Hydro ART code \citep{Kravtsov97, Kravtsov03, Ceverino09}.
 The ART code uses an adaptive mesh refinement approach to track the evolution of gravitating $N$-body and Eulerian gas dynamics. It also incorporates many relevant processes for galaxy formation like cooling by atomic and molecular hydrogen (also metals), star formation, and feedback from supernovae, stellar winds, and radiation. These processes are included as state-of-the-art subgrid models \citep{Ceverino17b}. 
 
 In short, star formation is assumed to occur at densities above 1 cm$^{-3}$ and at temperatures below $10^4$ K. The code asumes a stochastic star formation model that reproduces the empirical Kennicutt-Schmidt law \citep{Schmidt, Kennicutt1998}.
In addition to the injection of thermal energy, the simulations use radiative feedback, as a local approximation of radiation pressure. This non-thermal pressure is added to the total gas pressure in regions where ionizing photons from massive stars are produced and trapped. The model of radiative feedback is named RadPre\_IR in \cite{Ceverino14} and it uses a moderate trapping of infrared photons. 
Finally, the feedback model also includes the injection of momentum coming from the (unresolved) expansion of gaseous shells from supernovae and stellar winds \citep{OstrikerShetty11}.  More details can be found in \cite{Ceverino17b}, \citet{Ceverino09}, and \citet{Ceverino14}.  
 
The simulations follow the advection of metals released from SNe-Ia and from SNe-II, using supernovae yields that approximate the results from \cite{WoosleyWeaver95}, as described in \cite{Kravtsov03}.
 The standard simulations (10-Mpc and 20-Mpc boxes) have the following resolution: a dark matter particle mass resolution of $10^4 M_{\odot}$, a minimum star-particle mass of $100 M_\odot$, and a maximum spatial resolution of $8.7 - 17$ pc. 
The simulation of the 40-Mpc-box has 8 times lower mass resolution and twice lower spatial resolution.

\subsection{Sample of FL galaxies}

The FL suite covers a mass and volume complete sample of galaxies at $z=5$. It contains all halos with a maximum circular velocity higher than 50 km s$^{-1}$ for the 10-Mpc box, and 100 km s$^{-1}$ for the 20-Mpc box. 
The 40-Mpc-box adds another set of 50 simulated galaxies with circular velocities higher than 200 km s$^{-1}$.
Despite the small cosmological volumes, 
the resulting sample of 355 simulated galaxies \citep{Ceverino17b} covers a large range of halo masses from $M_{virial} \simeq 10^{9}$ to $10^{12} M_{\odot}$.

\begin{figure}
	\includegraphics[width=\columnwidth]{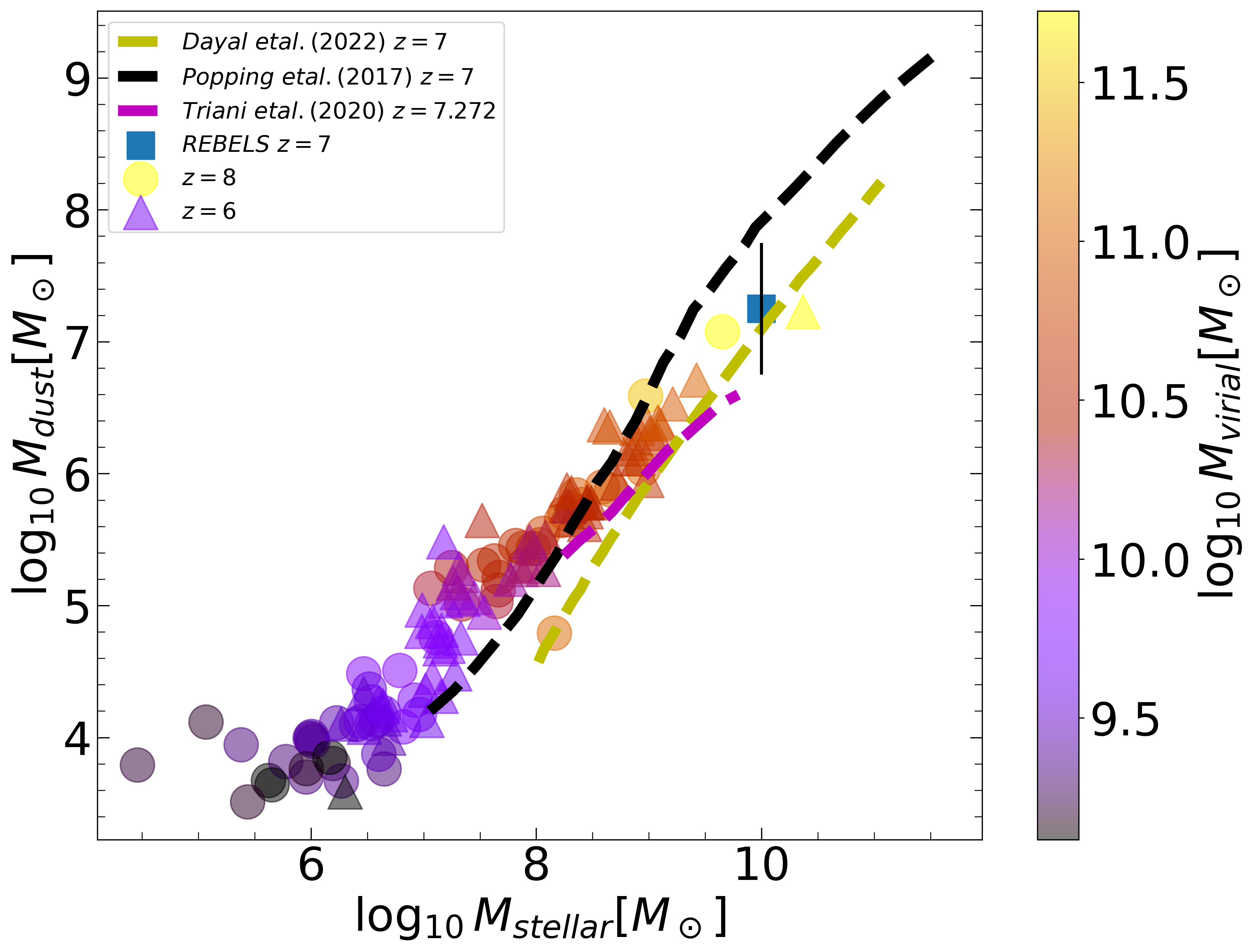}
    \caption{Relation between physical properties of 120 galaxies used in this work at $z=6$ (triangles) and $z=8$ (circles). The virial mass of the galaxy is indicated using colors. Here we also compare with the results from SAMs \citep{popping17, triani20, dayal22} as shown by Black, Yellow, and Magenta dash-lines respectively. The square symbol represents observations at $z\simeq7$ by the REBELS survey \citep{Bouwens22}. }
    \label{fig:one}
\end{figure}

According to the stellar mass function of FL galaxies \citep{2018DC}, our set covers a diverse halo mass range at different redshifts. We randomly select a subsample of 120 objects: 60 galaxies at $z=6$ and 60 galaxies at $z=8$, covering the full range of virial masses.
The relationship among physical properties is represented in Fig.~\ref{fig:one}. 
Here, the stellar mass of a galaxy is defined within a sphere of  $15\%$ of their virial radius, ${\rm R}_{\rm virial}$ \citep{BryanNorman}.
The dust mass is extracted from the relation,
\begin{equation}
    M_{\rm dust} [M_{\odot}] = \sum\limits_{i = 1}^N \rho_{\rm dust, i}.dV_i = \sum\limits_{i=1}^N (DMR\cdot\rho_{\rm gas,i}\cdot Z_i)\cdot dV_i \ .
	\label{eq:one}
\end{equation}
Here $\rho_{\rm dust,i}$ is the dust density in each cell where $i$ is the cell index, and $dV_i$ is the corresponding volume of the cell. Dust density is calculated from the gas metallicity, $Z_i={\rm M}_{\rm metal,i}/{\rm M_{\rm gas,i}}$, and the gas density within each cell. All the cells within $15\%$ of the virial radius are considered.
$DMR$ is the  dust-to-metal ratio and the actual value is very uncertain. It depends on the balance between dust formation and destruction. These scenarios become more important for hotter environments near stars and they may vary  with redshift. For simplicity in this first study, we adopt a constant value of $DMR = 0.4$, which is a good approximation for high-$z$ galaxies \citep{2021DMR}. We note in this context that \cite{2019dust} explored different values of $DMR=0.2, 0.4$, and 0.8 and got similar dust temperatures and IR luminosities.

The sample covers a large range of galaxy masses, from  ${\rm M}_{\rm stars} \simeq 10^{5}$ to $10^{10} M_{\odot}$ and consequently the range of dust masses is also large, ${\rm M}_{\rm dust} \simeq 10^{4}$ to $10^{7} M_{\odot}$. This means that we sample different galaxy conditions. The dust-to-stellar ratio is not constant due to the different metallicities of these galaxies \citep{2020langan}. 
We compared with previous works based on semi-analytic models (SAMs) of galaxy formation \citep{popping17, triani20, dayal22}, which track the production and destruction of dust within galaxies at cosmic dawn.
We found a reasonable agreement within a common mass range.
Although these models have variable $DMR$, they show similar results within a factor of a few. Therefore, the overall impact of a variable $DMR$ on the dust-stars mixture is relatively small within this mass range.
The most massive galaxies in the FL sample are analogs to observed galaxies from the Reionization Era Bright Emission Line Survey (REBELS) program \citep{Bouwens22}.

\subsection{Extracting SEDs of star-particles using BPASS}

In order to handle the intrinsic radiation of star-particles, the binary population and spectral synthesis model \cite[BPASS\footnote{\href{https://bpass.auckland.ac.nz/index.html}{https://bpass.auckland.ac.nz/index.html}}, see ][]{Eldridge17} is used. This model generates the rest-frame UV and optical SEDs for each star-particle according to its age, metallicity, and mass of a single stellar population (SSP). The wavelength range extends from 1 to $10^5$ \AA. 
BPASS assumes a Kroupa initial mass function \citep{Kroupa:2002ky} with power slopes $-1.3$ and $-2.35$ for a star mass-range of $0.1-0.5 M_{\odot}$ and $0.5-100 M_{\odot}$ respectively. BPASS uses a grid of 13 values of metallicity ($Z=10^{-5}-0.04$), and 40 logarithmic bins in SSP ages between 1 Myr and 100 Gyr. A detailed analysis of the intrinsic spectra of FirstLight galaxies shows a peak emission in the FUV  due to the fact that these high-$z$ galaxies are star-forming and metal-poor \citep{2019DC}. 

A caveat in our procedure is related to the large numbers of star-particles within the most massive halos.
It would be computationally expensive to run an MC-RT simulation using every star-particle in these cases. In order to overcome this problem for massive halos in which the number of star-particles within the galaxy exceeds $10^4$, we select $10^4$ random star-particles and increase their SEDs by the corresponding factor, $\text{N}_{\mathrm{s}}/ 10^4$, where $\text{N}_{\mathrm{s}}$ is the total number of star-particles. In order to test if the total and selected star-particles show the same distribution, we  applied the Kolmogorov-Smirnov method. It  shows that the deviation between population and sample data is quite low, verifying our null hypothesis for all galaxy masses \citep{ourpaper}.

\subsection{POLARIS RT simulator}

\begin{figure}
	\includegraphics[width=\columnwidth]{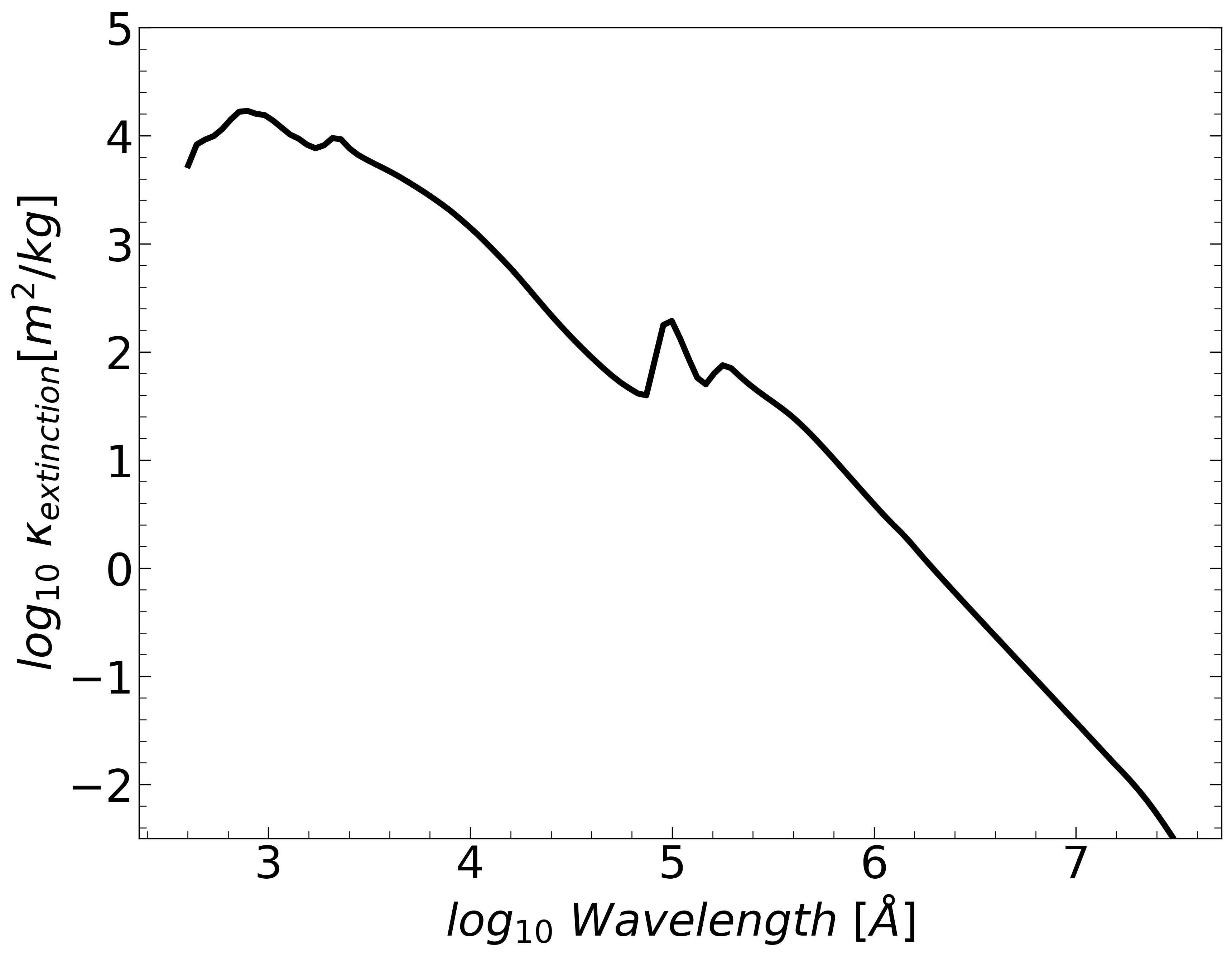}
	\caption{Dust extinction used in POLARIS, considering both scattering and absorption of silicate and graphite dust grains.}
    	\label{fig:two}
\end{figure}

Once we know the intrinsic spectra of star-particles and the spatial distribution of stars and dust within a galaxy, we need a dust model as a representation of the optical properties of dust grains. 
Spherical graphite and silicate compositions have been considered with the material densities $\rho = 2250$ and 3800 $\rm kg/m^3$, and a mass fraction of 0.625 and 0.375, respectively. 
For this composition, the grain-size dust model of \cite{1977ApJ...217..425M} has been implemented.
This gives a power-law distribution, $n_{d}(a) = a^{-3.5}$, in the range of $5 \times 10^{-3} {\rm \mu m} < a < 0.25 {\rm \mu m}$; where $a$ is the grain size. The selection of grain-size distribution describes the dimensionless absorption and scattering efficiencies, $Q_{abs} = C_{abs}/\pi a^2$ and $Q_{sca} = C_{sca}/\pi a^2$ respectively.
Here $C_{abs}$ and $C_{sca}$ are the cross-section of absorption and scattering. 
The optical properties of grains are extracted from \citet{2001_weingartner}. 
The resulting dust opacity as a function of wavelength is shown in Fig. \ref{fig:two}.
As expected, the opacity at shorter wavelengths is higher. 

After assembling all inputs, we run two simulations with POLARIS: (1) CMD-TEMP and (2) CMD-DUST-EMISSION. The former calculates the dust temperature of ISM in a galaxy via the physical process of continuous absorption and immediate re-emission \citep{1999lucy}. The main assumptions are that the dust particles are in local thermodynamic equilibrium (LTE) and dust is the only opacity source. 
In radiative equilibrium, the absorbed photon energy must be re-emitted instantly at a longer wavelength that is simply expressed by the Planck function $B(\nu, T)$ which is a function of frequency and temperature \citep{2001bjorkman,stefan},
\begin{equation}
    \frac{dP_{i}}{d\nu} = \frac{\kappa_{\nu}}{K} \left(\frac{dB(\nu, T)}{dT}\right)_{T=T_i} \ ; \ \  \ \ \ \ \ \ K = \int_0^\infty \kappa_{\nu} \left(\frac{dB(\nu, T)}{dT}\right) d\nu .
	\label{eq:three}
\end{equation}
Here $\frac{dP_{i}}{d\nu}$ is the probability of re-emitting the photon between frequencies $\nu$ and $\nu+d\nu$ with $i$ being the cell index, and $K$ is the normalization constant. We do not consider any process of non-equilibrium heating of dust that is highly time-dependent and varies with dust size (e.g stochastical heating). 

Once the dust grains achieve a certain temperature distribution, the second simulation in POLARIS is used to calculate the rest-frame, transmitted radiation field. The UV-optical and thermal mid-far IR light is detected by an imaginary ‘plane detector\footnote{The characteristics of the plane detector are: observing distance = $3.086 \times 10^{25} m$, viewing angles $(x=y=0)$,  wavelength range between $3 \times 10^2$ and $3\times 10^{7}$ \AA, divided into 120 bins}’, which is appropriate for light coming from very far objects so that the incoming light rays are considered parallel. 

POLARIS treats light as a package of photons having specific frequencies emitted from the sources which select random paths from one scattering event to the next, as it passes through the gas and dust cloud. In this process, the photon frequency remains the same, but the angle gets redistributed according to the Henyey-Greenstein phase function. This phase function depends on the asymmetry parameter that controls the distribution of scattered light \citep{10.5555/3044800}. If it is an absorption+re-emission event then the new frequency and direction are chosen randomly. 

\begin{figure}
	\includegraphics[trim={0 2 0 0},clip, width=\columnwidth]{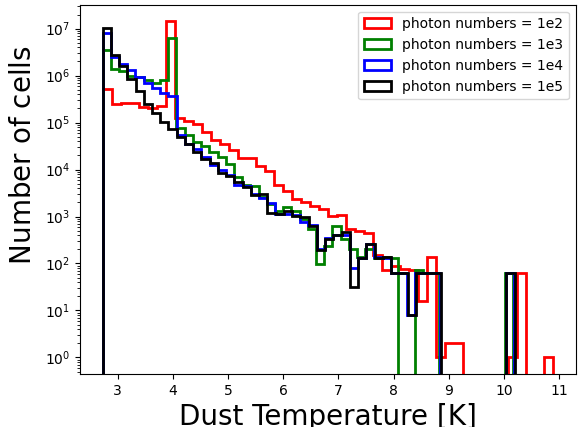}
    \caption{Testing different number of photons per wavelength $(N_{ph}/\lambda)$ on a simulated galaxy. The distribution of dust temperatures reaches convergence at $N_{ph}=1000$ per wavelength bin.}
    \label{fig:three}
\end{figure}

POLARIS requires as input the number of photons per wavelength that is radiated from the stars. Due to the random selection of photon directions, if the number of photons is small, it may generate noise that affects the final simulated results. The noise is inversely related to the square root of photon numbers $(\propto 1/\sqrt{N_{ph}})$. Usually, one million photons are used but this can be time-consuming for some galaxies. In order to find a reasonable number of photons, $N_{ph}$ per wavelength bin, that is sufficient for this purpose, we tested a sample galaxy and simulated its dust temperature by using different numbers of photons. From this analysis, we conclude that $N_{ph}=1000$ gives a robust result, as shown in Fig. \ref{fig:three}, where the distribution of dust temperatures has converged above that number.

\section{Results and Discussion}
\label{sec:resultsdis}
\subsection{Dust attenuation curve}

\begin{figure*}
	\includegraphics[width=1.8\columnwidth]{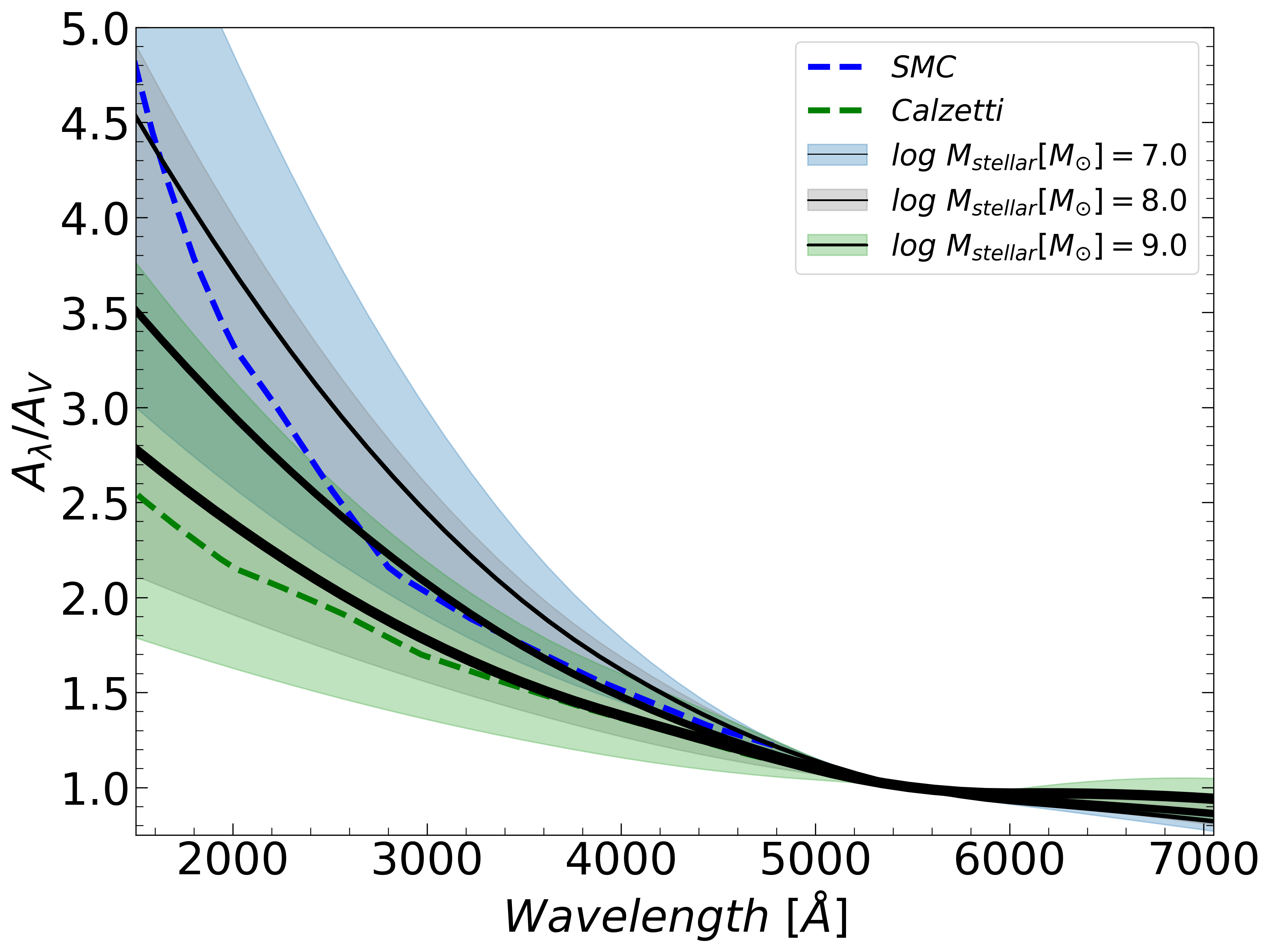}
    \caption{Attenuation curves for galaxies in three 1-dex mass bins (solid lines). Shaded areas show the $1\sigma$ deviation around the averages. The low-mass galaxies have attenuation curves consistent with the SMC curve (blue dash line). Higher masses show shallower curves, mostly consistent with the Calzetti attenuation curve (green dash line).}
    \label{fig:four}
\end{figure*}

The intrinsic light from stars is attenuated by the intervening dust.
The transmitted light includes the emission from unobscured stars, the 
fraction of light scattered back into line-of-sight, and the fraction that is not lost by absorption.
Attenuation of the galaxies is evaluated along random line-of-sights. Hence, it can include the effects arising from the distribution of stars and dust within the galaxy. 
This complex distribution depends on the galaxy's mass. 
For example,  galaxies with high stellar mass generally have much denser regions of dust. 
To analyze these effects, we compute the dust attenuation $A_\lambda$ as a function of  wavelength \citep[e.g.][]{2020salim}, 
\begin{equation}
    A_{\lambda} = m_{\lambda} - m_{\lambda,0}= 2.5 \times \log_{10}\left(L_{\lambda,0}\right) -2.5 \times \log_{10}\left(L_{\lambda}\right)
    \label{eq:four}
\end{equation}
where $m_{\lambda,0}$ ($L_{\lambda,0}$) is the intrinsic, unattenuated, monochromatic magnitude (luminosity) and $m_{\lambda}$ ($L_{\lambda}$) is the attenuated/transmitted magnitude (luminosity).
The wavelength range of interest ranges from 1500 to 7000 \AA. 

In principle, the behaviour of attenuation is directly related to the dust optical depth and the dust column density within galaxies. 
Massive galaxies at high-$z$ have high dust masses (Fig.~\ref{fig:one}) and therefore we expect that their attenuation is higher than in low-mass galaxies. However, dust attenuation is a complex interplay between dust and starlight. It behaves differently in different galaxies even with the same stellar mass due to geometrical effects \citep[e.g.][]{2020salim}.
Therefore, we study the normalization and the shape of the attenuation curve separately. The shape can be described by $A_\lambda/A_V$, where $A_V$ is the attenuation at the V-band ($\lambda = 5500$ \AA). There are diverse techniques that define the normalization in different ways, such as E(B-V) or optical depth \citep[e.g.][]{refId0, Battisti_2017, Narayanan_2018}. Here we use the rest-frame FUV attenuation, $A_{1500}$. This wavelength is relevant in high-$z$ observations, because (a) it is bright in star-forming galaxies at high-$z$, (b) it has no absorption lines, and (c) dust, if present, preferentially attenuates this FUV regime.

Figure \ref{fig:four} shows different attenuation curves in different, 1-dex mass-bins.
The wide shaded areas indicate the standard deviation $(1\sigma)$ above and below the averaged attenuation curves of all galaxies at $z=6, 8$, binned by their stellar masses. 
The general feature of all these curves is 
the steep rise from the visible to the far-UV. 
Although it is a non-linear curve, it can be approximated by a power-law with an exponent, $-n$, in the range between 1500 and  5500 \AA.
Attenuation curves can significantly depart from this single power-law but it is useful to characterize its steepness or shallowness. 
Different empirical models have been proposed based on the properties of different galaxies, such as the Small Magellanic Cloud (SMC), see \citep{1984prevot} or local starburst galaxies \citep{1994ApJ...429..582C}. 
Both curves are also shown in Fig. \ref{fig:four}. The SMC attenuation curve is steeper in FUV ($n=1.20$) than the Calzetti law ($n=0.70$). In general, the steepness is associated with the geometry of the stars-dust mixture and radiative transfer effects like clumpiness that can affect the attenuation curves. Massive galaxies are considered to be more clumpy, have high metallicity, and are optically thick, resulting in shallower attenuation curves than in low-mass galaxies \citep{Shivaei_2020}. 

The shape of the attenuation curve depends on the galaxy's stellar mass. Curves in low-mass galaxies, $M_{\rm stars}\simeq 10^7 - 10^8 M_{\odot}$ are closer to the SMC attenuation curve. On the other hand, more massive galaxies, $M_{\rm stars}\simeq 10^9 M_{\odot}$, have shallower curves, similar to the Calzetti law. 
The attenuation curve of the most massive bin, $M_{\rm stars}\simeq 10^{10} M_{\odot}$, is not shown because that mass bin only contains two galaxies.
The dust attenuation law of a given galaxy evolves with time as the galaxy grows.
Galaxies at high-$z$ transition from a regime with low dust column density, low metallicities and an extinction curve similar to SMC to a regime with high dust column density and shallower attenuation curves at lower redshifts. 
Therefore, as the mass and metallicity of a galaxy increase, the slope of the attenuation curve becomes shallower.

The variations between galaxies depend on the morphology of the spatial distribution of stars and dust. \cite{2018Corre} compared the attenuation curves obtained with the CIGALE code \citep{cigale} with an RT simulator introduced by \cite{2016ApJ...833..201S}. CIGALE integrates a stellar spectral energy distribution (SED) in the UV-optical range with a dust component that radiates in the IR spectrum. The code ensures that there is a complete conservation of energy balance between the dust-absorbed radiation and its subsequent re-emission in the IR range. \cite{2018Corre} found that shallow curves are obtained for a high optical depth and a very clumpy ISM. In addition, steep curves are the results of a low optical depth along with a homogeneous or shell-like ISM. 
According to \cite{10.1093/mnras/stt523}, steep curves are driven by the dominance of scattering over absorption and the fact that scattering is more forward-directed at shorter wavelengths whereas it is more isotropic at longer wavelengths. Shallow curves experience more absorption than scattering due to the high optical depth. 

\subsubsection{Parameterization  of attenuation curves}
\label{sec:parameterization}

We now compare the attenuation curves described above with other curves commonly used in high-$z$ galaxies. For example, \cite{2000calzetti}, Cal00 hereafter, proposed a relatively gentle slope. It is described in two wavelength ranges: (a) a third-order polynomial fit in $\lambda^{-1}$ for the UV/blue optical regions, and (b) a linear fit in $\lambda^{-1}$ for red optical/near-IR regions, as follows
\begin{equation}
\begin{split}
  K_{\rm Cal,\lambda} &= \frac{A_{\lambda}}{A_{V}} R_{V} 
  =\begin{cases}
    2.66\times a(\lambda)+R_V, \ \text{$0.12\le \lambda \le 0.63\mu {\rm m}$}\\
    2.66\times b(\lambda)+R_V, \ \text{$0.63<\lambda \le 2.20\mu {\rm m}$} \ .
  \end{cases}
\end{split}
\end{equation}
Here $a(\lambda)=(-2.156 + 1.509/\lambda - 0.198/\lambda^{2} + 0.011/\lambda^{3})$ and $b(\lambda)=(-1.875 + 1.040/\lambda)$, additionally $R_{V}$ describes the correlation between the shape of UV extinction with the regions of IR-Optical that is shown by $R_V = A_V / E(B-V)$, where $E(B-V)$ is the color excess that represents reddening effect. It is expressed as, $E(B-V) = (B-V)_{\text{attenuated}} - (B-V)_{\text{unattenuated}}$. 

There are other functional forms of attenuation curves, like the modified Calzetti model that separately includes a Drude function, $D_{\lambda}(\mathcal{B})$, and a power-law term $\delta$. The Drude function describes the UV bump with amplitude $\mathcal{B}$ at the fixed central wavelength (2175 \AA) and width ( 350 \AA). This version is widely used in SED fitting models like CIGALE \citep{cigale}. It is expressed as,
\begin{align}
    K_{\rm ModCal,\lambda} &= K_{\rm Cal,\lambda} \frac{R_{V, \rm mod}}{R_V} \left(\frac{\lambda}{5500}\right)^{\delta} + D_{\lambda}(\mathcal{B}) \\
    D_{\lambda}(\mathcal{B}) &= \frac{\mathcal{B} \lambda^{2}(0.35 \mu m)^{2}}{[\lambda^{2}-(0.2175 \mu m)^{2}]^{2} + \lambda^{2}(0.35 \mu m)^{2}}
	\label{eq:drude}
\end{align}
The relationship between $R_{V, \rm mod}$ and $\delta$ can be derived by imposing $E(B-V)=1$ \citep{2018salim}:
\begin{equation}
    R_{V,mod} = \frac{R_V}{(R_V+1)(4400/5500)^{\delta}-R_V} \ ,
	\label{eq:eight}
\end{equation}
The modified curve can be reduced to Cal00 by setting $D_{\lambda}(\mathcal{B}) = \delta = 0$. Normally, the ranges of $D_{\lambda}(\mathcal{B})$ and $\delta$ depend on the stellar masses of the star-forming galaxies as shown by \cite{2018salim}. The steep far-UV rise of an attenuation curve is non-linear between UV and the optical, but to some extent, we can define it approximately by the power law between extinction at 1500 \AA \ and in the V-band as, 
\begin{equation}
    \frac{A_{\lambda}}{A_{V}} = \left(\frac{\lambda}{5500}\right)^{-n}\ .
	\label{eq:seven}
\end{equation}

For $\lambda = 1500$ \AA, the power-law index can be described as $n$ reduces to $n = 1.772 \times \log_{10} (A_{1500}/A_V)$ \citep{2020salim}. 

\begin{table}
	\centering
	\caption{Values of the multi-regression model of attenuation, \equ{nine}, for $z = 6, 8$, based on stellar mass and absolute magnitude.}
	\label{tab:table2}
	\begin{tabular}{lcccr} 
		\hline
		$z$ & $m_1$ & $m_2$ & $\mathcal{K}$ & $R^2$\\
		\hline
		6.0 & 0.0319 & -0.3083 & -6.8107 & 0.882\\
		8.0 & 0.5585 & -0.0953 & -6.9932 & 0.816\\
		\hline
	\end{tabular}
\end{table}

\begin{figure}
\includegraphics[width=\columnwidth]{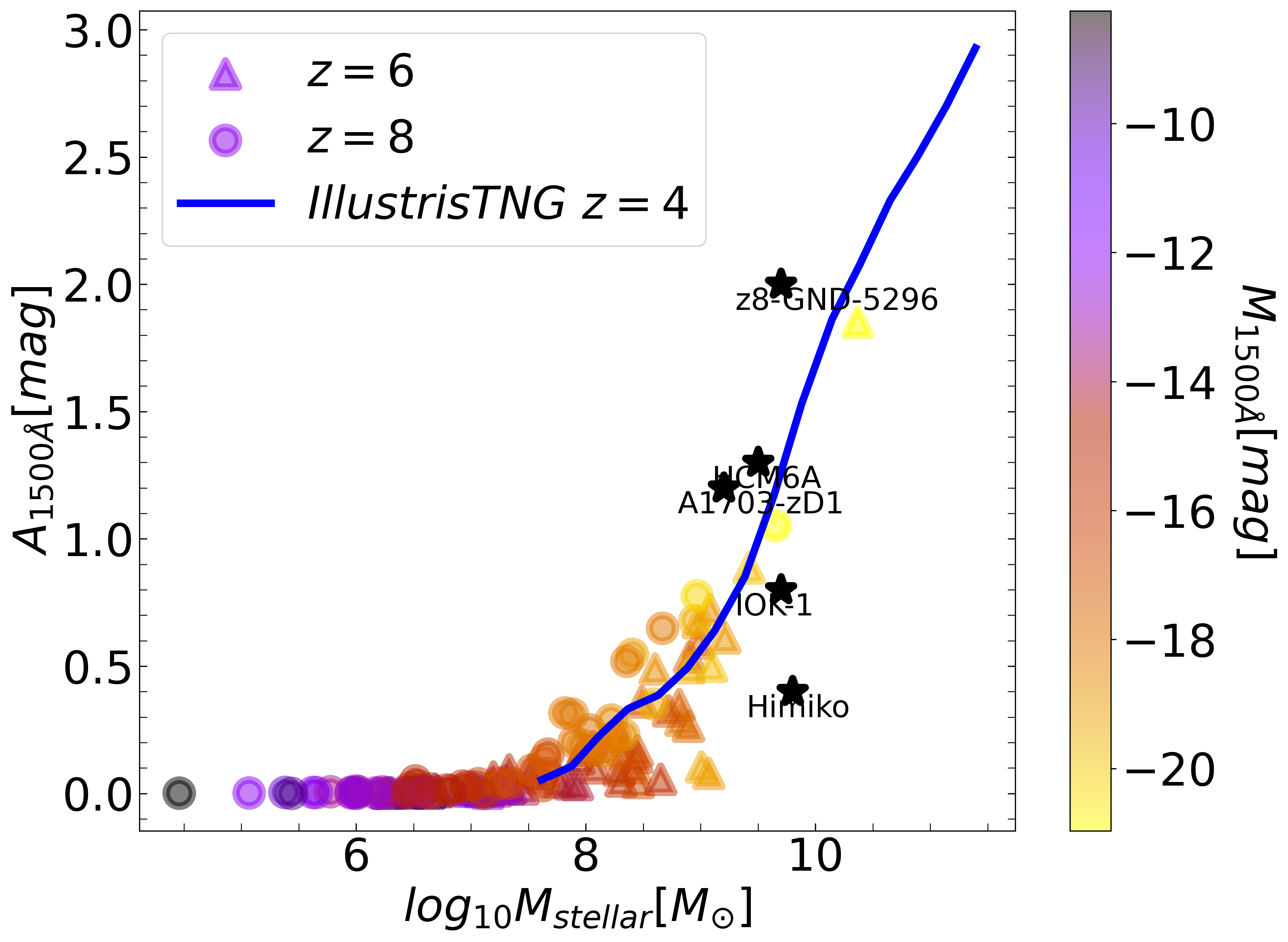} 
\includegraphics[width=\columnwidth]{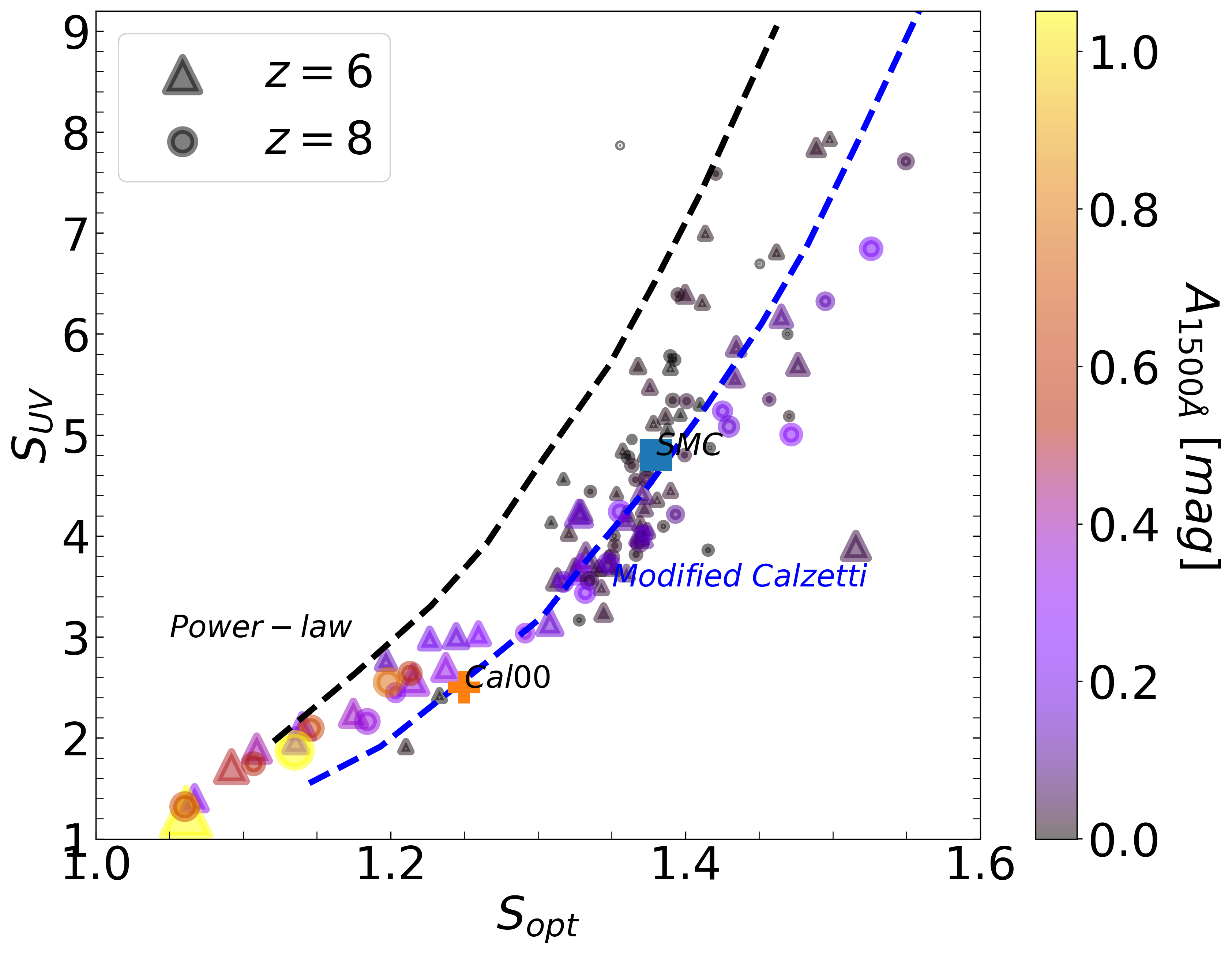}
\caption{Attenuation curves parametrized by FUV attenuation at 1500 \AA \ (top). UV slope ($A_{1500}/A_V$) and optical slope ($A_B/A_V$) are shown in the bottom panel. Galaxies are represented by triangles and circles for $z = 6$ and 8 with color bars showing absolute magnitude and attenuation at 1500 \AA, and the size of symbols represents the galaxy stellar mass. We compare with the attenuation models of the power-law curve ($0.5<n<1.7$), Modified Calzetti curve  ($-1.0<\delta<0.4$ ), SMC, and Cal00. Black stars are high-$z$ observations \citep{observations2015} and  the solid blue line represents the illustrisTNG simulation \citep{10.1093/mnras/staa1423}.}
\label{fig:five}
\end{figure}

\begin{figure*}
	\includegraphics[width=1.8\columnwidth]{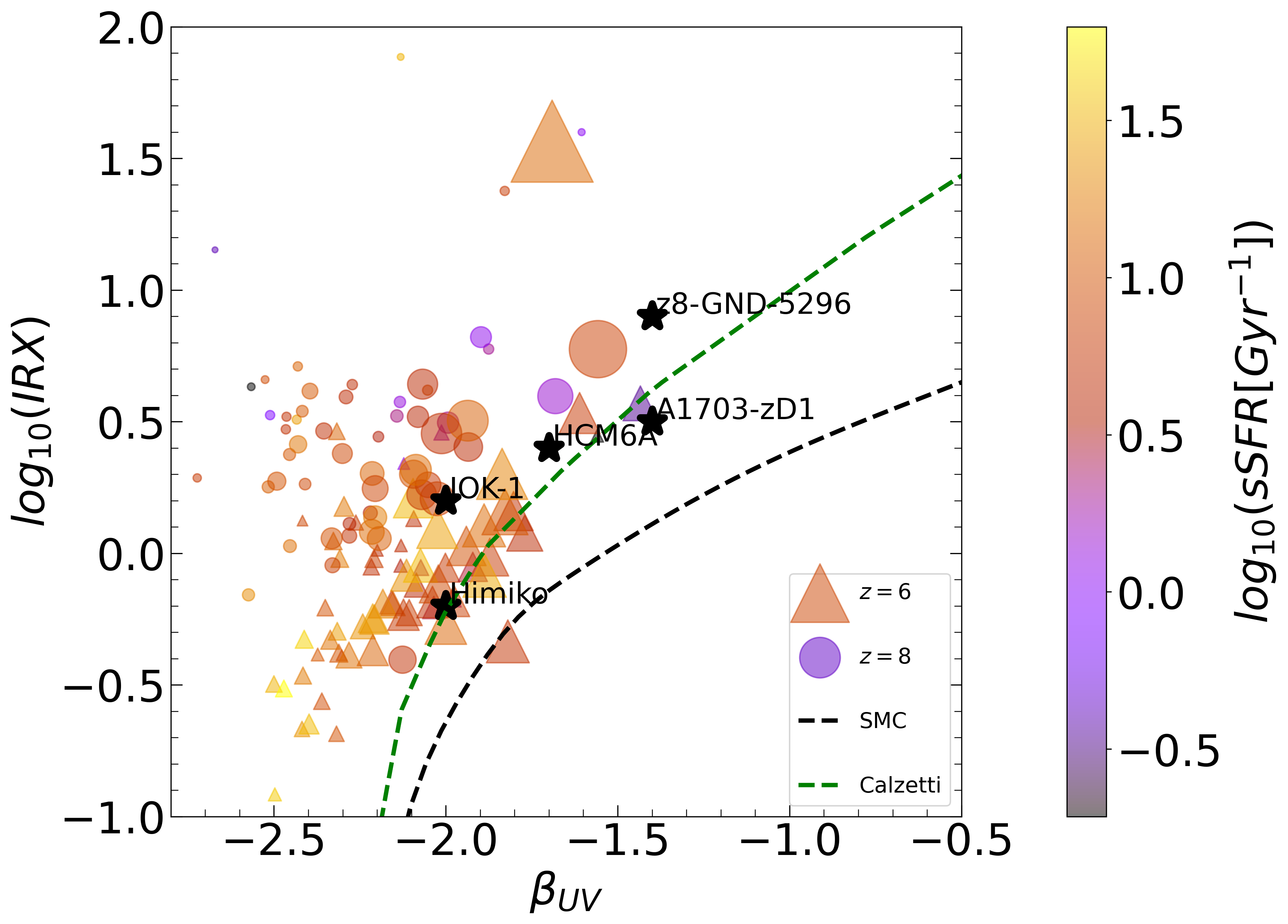}
    \caption{\textbf{$IRX-\beta_{UV}$} relation for z= 6, 8 shown by triangles and circles. The  colormap represents the sSFR. The size of the symbols represents the stellar masses. Here we also compare with the high-$z$ observations \citep{observations2015}, Calzetti \citep{2000calzetti}, and SMC models \citep{1984prevot}.}
    \label{fig:six}
\end{figure*}

Exploring  attenuation curves requires high spectral resolution.
Here we are mostly interested in the overall shapes of these attenuation curves in the UV and optical bands. Therefore, we define the UV slope of the attenuation curve as
\begin{equation}
    S_{\rm UV}=A_{1500}/A_V
\end{equation}
where $A_{1500}$ is the attenuation at 1500 \AA.
The optical slope is defined as
\begin{equation}
    S_{\rm opt}=A_B/A_V
\end{equation}
Figure \ref{fig:five} illustrates the relation among the different parameters as defined above. 
Each point represents an individual galaxy in our FirstLight samples at $z=6$ and 8.
The top panel shows the attenuation at $\lambda=1500$ \AA \ versus stellar mass. 
Values of $A_{1500}$ near zero indicate that 100$\%$ of the UV light at 1500 \AA \ escapes (no dust attenuation) and higher values represent more dust absorption. 
We find that at any redshift, lighter galaxies are optically thin, and more massive galaxies have a higher attenuation and they are more opaque to UV light. 
We conclude that the lower mass limit required to absorb a significant amount of UV light ($A_{1500}>0.1$) is about $10^8 M_{\odot}$.
This limit corresponds to an absolute magnitude of $M_{1500} \simeq -16$ \citep{2019DC}. Since the intrinsic properties of galaxies are interconnected with the degree of attenuation, we applied a multiregression model for $A_{1500}$ based on  $M_{\rm stars}$ and $M_{1500}$. It is expressed as,
\begin{equation}
    \log_{10} (A_{1500} [{\rm mag}]) =  m_1\cdot \log_{10} (M_{\rm stars} [M_{\odot}]) + m_2\cdot M_{1500} [{\rm mag}] + \mathcal{K}
	\label{eq:nine}
\end{equation}
Here $m_1, m_2$ are the coefficients and constant $\mathcal{K}$ is the $y$-intercept. Their values for $z=6, 8$ are shown in Table. \ref{tab:table2}. 

Observations of galaxies at similar redshifts show hints of this relation between stellar mass and UV attenuation.
We compare with observations of galaxies at $z>6$ \citep{observations2015}. They estimate FUV attenuation at $\simeq1800$ \AA \ by using the total IR to total UV luminosity ratio. 
Their estimates of FUV attenuation agree with our trends. 
We also compare with the results of the IllustrisTNG simulation at $z=4$ \citep{10.1093/mnras/staa1423}. They also found a tight relation between $A_{UV}$ and log($M_{\rm stars}$). 
The relation is linear for most masses but it becomes flatter at low $M_{\rm stars}$, which is consistent with our findings. 

The bottom panel of Fig. \ref{fig:five} shows the relation between UV slope and optical slope. We compare the results from FirstLight with power-law and modified Calzetti, Cal00, and SMC attenuation models. 
The general picture is that FirstLight galaxies agree well with the modified Calzetti model. 
Low-mass galaxies tend to have negative $\delta$ values and their curves are closer or even steeper than the SMC curve.
On the other hand, more massive galaxies are consistent with positive values and they have curves similar to Calzetti, as seen in \fig{four}. 
There is a slightly affinity towards power-law attenuation curves for high masses and highly attenuated galaxies \citep{2020salim}. 
Their curves show a gentle rise in both UV and optical bands 

\subsection{IRX vs $\beta_{UV}$}

The empirical relation between the infrared excess, $IRX=L_{IR}/L_{UV}$, and the reddening of the UV color, measured by the $\beta_{UV}$ slope, provides some hints of the complex interplay between dust and UV light in star-forming galaxies  \citep{Meurer1995}. As dust attenuation increases, the UV slope flattens, and the infrared excess increases \citep{Meurer1999}. However, there is large scatter in this relation \citep{scatterIRXbeta1}. 

We illustrate the $IRX-\beta_{UV}$ relation for the simulated galaxies at $z=6$, 8, and the comparison with high-$z$ observed galaxies \citep[][]{1984prevot, 2000calzetti, observations2015} in Fig.~\ref{fig:six}. 
In general, FirstLight is closer to the Calzetti prescription, although there is a general shift towards lower $\beta_{UV}$ values, especially at low masses. 
The origin of such relation lies in the intrinsic properties of high-$z$ galaxies. 
Low-mass galaxies have lower $\beta_{UV}$ and $IRX$ values.
This is due to low metallicities, young stellar ages and low UV attenuation. As a result, the transmitted light from galaxies is very similar to the intrinsic light, which varies between $\beta_{UV}\simeq-2.3$ and -2.4 on average \citep{2019DC}. These slopes are steeper than in galaxies with a similar infrared excess at lower redshifts \citep{2020salim}.
For massive galaxies, dust column densities are much higher. This increases the absorption of UV photons, which flattens the UV spectrum and increases the infrared excess \citep{2017narayan}. 
This trend is consistent with observations \citep{observations2015} that follow the Calzetti model with a significant scatter.

At $z=8$, galaxies experience a shift to higher $IRX$ values. The total IR luminosity of a galaxy at $z$=8 is higher than at $z$=6 for similar dust reddening. This is because of the stronger CMB radiation at higher redshifts. It heats dust to higher temperatures. As a result, there is more IR emission at higher redshifts \citep[e.g.][]{2013cunha} at a fixed $\beta_{UV}$ value.   

Some FirstLight galaxies do not follow  the general trend of the sample, because of their relatively high infrared excess, $IRX>10$.
These high $IRX$ values are the combined result of dust geometry and CMB heating. In these galaxies, their dust structure is very complex and spread out across several kpc. This dust is mainly heated by the CMB radiation \citep{ourpaper}. 
This is particularly efficient for low-mass galaxies. 
Massive galaxies are mainly heated by the strong UV radiation from stars. 
The most massive galaxy at $z=6$ has the highest attenuation, $A_{1500}=2$.
It also has a very high infrared excess, $IRX\simeq30$, but it has a relatively steep UV slope, $\beta_{UV}\simeq-1.7$, inconsistent with Calzetti. 
This is the result of the relatively flat attenuation curve, $S_{\rm UV}\simeq1$  of this particular galaxy. 
Another galaxy of a similar mass at $z=8$ has a lower \textit{IRX}, more consistent with Calzetti
because its attenuation curve is steeper,  $S_{\rm UV}\simeq2$.
More examples of massive galaxies are needed to address the relevance of outliers.

\section{Conclusion}\label{sec:conclusion}

The attenuation of dust at cosmic dawn is studied by using FirstLight (FL) zoom-in cosmological simulations at $z = 6$ and 8, with the help of the POLARIS RT simulator. 
The main inputs required to simulate the radiative transfer process are: (1) the distribution of star-particles and dust within the galaxies, extracted from FL, (2) the intrinsic stellar spectra from the BPASS model according to the ages, masses and metallicities of the FL star-particles, (3) optical properties of dust grains, which rely on the grain-size distribution and dust extinction. 

The effect of attenuation by dust is described through the attenuation curves and the relation between the UV slope, $\beta_{UV}$, and the infrared excess, $IRX$. Here are some key results of our work: 
\begin{enumerate}
    \item The attenuation curves follow the Calzetti dust model only for relatively massive galaxies, $M_{\rm stars}\simeq 10^9 M_{\odot}$. Galaxies with lower masses have steeper curves, consistent with the SMC model (Fig. \ref{fig:four}).
    \item The attenuation at 1500 \AA \ depends strongly on the stellar masses and the absolute magnitude of galaxies at $z=6$ and 8, consistent with observations and other simulations (Fig. \ref{fig:five}). 
    \item The UV and optical slopes of the attenuation curves are consistent with the modified Calzetti model, with a slightly preference to the power-law model for the galaxies with the highest values of attenuation (Fig. \ref{fig:five}).
    \item The $IRX-\beta_{UV}$ relation at $z=6$ follows the Calzetti model with a shift to slightly lower $\beta_{UV}$ values due to lower metallicities (Fig. \ref{fig:six}).
    \item The same relation at $z=8$ shows a shift to higher \textit{IRX} values due to a stronger CMB radiation at high-$z$.
\end{enumerate}

In this pilot study, we have addressed the diversity of attenuation curves in galaxies at cosmic dawn. However, there are many more open questions that we aim to answer in future work:
what is the dust-mass growth rate in galaxies and how it affects the attenuation? 
How diverse is the attenuation at high masses $M_{\rm stars} \ge 10^{10} M_{\odot}$?

\section*{Acknowledgement}
The authors are grateful to the IWR Heidelberg Computer Server to allowed us to simulate and store heavy simulation data. The authors gratefully acknowledge the Gauss Center for Supercomputing for funding this project by providing computing time on the GCS Supercomputer SuperMUC at Leibniz Supercomputing Centre (Project ID: pr92za).
The author thankfully acknowledges the computer resources at MareNostrum and the technical support provided by the Barcelona Supercomputing Center (RES-AECT-2020-3-0019).
We thank the BPASS team for sharing their database of SSPs and emission lines. This work made use of the v2.1 of the Binary Population and Spectral Synthesis (BPASS) models as last described in \cite{Eldridge17}. 
DC is a Ramon-Cajal Researcher and is supported by the Ministerio de Ciencia, Innovaci\'{o}n y Universidades (MICIU/FEDER) under research grant PID2021-122603NB-C21. 
P.H.P. thanks FAPES for the financial support during the last stages of the project. 
This work was partly supported by the European Research Council via the ERC Synergy Grant ``ECOGAL'' (project ID 855130), by the German Excellence Strategy via the Heidelberg Cluster of Excellence (EXC 2181 - 390900948) ``STRUCTURES'', and by the German Ministry for Economic Affairs and Climate Action in project ``MAINN'' (funding ID 50OO2206). The authors also thank for computing resources provided by the Ministry of Science, Research and the Arts (MWK) of the State of Baden-W\"{u}rttemberg through bwHPC and DFG through grant INST 35/1134-1 FUGG and for data storage at SDS@hd through grant INST 35/1314-1 FUGG.

\section*{Data Availability}

The data underlying this article are available in the FirstLight database, at \url{http://odin.ft.uam.es/FirstLight} or
will be shared on reasonable request to the corresponding author.



\bibliographystyle{mnras}
\bibliography{main_v2} 





\bsp	
\label{lastpage}
\end{document}